\documentclass[letter]{aa} % for the letters 
\usepackage{graphicx,natbib}
\usepackage{txfonts}
\usepackage{ulem}
\newcommand{\Msun}{\ensuremath{M_{\odot}}}
\newcommand{\Rsun}{\ensuremath{R_{\odot}}}
\newcommand{\sh}{\ensuremath{\mathrm{sh}}}

\newcommand{\lc}{\ensuremath{\mathrm{lc}}}
\newcommand{\di}{\ensuremath{\mathrm{d}}}

\begin{document} 

   \authorrunning{Moriya, Sanyal, \& Langer}
   \titlerunning{Shock breakout in inflated stellar envelopes}

   \title{
Extended supernova shock breakout signals \\
from inflated stellar envelopes
 }

   \subtitle{}

   \author{Takashi J. Moriya
          \and
          Debashis Sanyal
          \and
          Norbert Langer
          }

   \institute{
              Argelander Institute for Astronomy, University of Bonn,
              Auf dem H\"ugel 71, D-53121 Bonn, Germany
             \\
              \email{moriyatk@astro.uni-bonn.de}
             }

   \date{Received 2014 September 5 / Accepted 2015 February 11}
%   \date{}

\abstract{
Stars close to the Eddington luminosity can have large low-density inflated envelopes.
We show that the rise times of shock breakout signals from supernovae can be
extended significantly if supernova progenitors have an inflated stellar envelope.
If the shock breakout occurs in inflated envelopes,
the shock breakout signals diffuse in them, and
their rise time can be significantly extended.
Then, the rise times of the shock breakout signals are dominated by
the diffusion time in the inflated envelope rather than the
light-crossing time of the progenitors.
We show that our inflated Wolf-Rayet star models whose radii are 
on the order of the solar radius can have shock breakout signals
that are longer than $\sim 100$ sec.
The existence of inflated envelopes in Wolf-Rayet supernova
progenitors may be related to the mysterious long shock breakout
signal observed in Type~Ib SN~2008D.
Extended shock breakout signals may provide evidence for the existence of
inflated stellar envelopes and
can be used to constrain the physical properties of these enigmatic structures.
}

% \abstract{}{}{}{}{} 
% 5 {} token are mandatory
 
%  \abstract
%  % context heading (optional)
%  % {} leave it empty if necessary  
%   {context}
%  % aims heading (mandatory)
%   {aims}
%  % methods heading (mandatory)
%   {methods}
%  % results heading (mandatory)
%   {results}
%  % conclusions heading (optional), leave it empty if necessary 
%   {conclusions}


   \maketitle
%
%________________________________________________________________

\section{Introduction}\label{introduction}
The first electromagnetic signals from a supernova (SN)
are emitted when the shock wave that travels through the stellar interior
reaches the stellar surface \citep[e.g.,][]{colgate1974,klein1978,ensman1992,matzner1999,tominaga2011,tolstov2013}.
Photons trapped in the shock wave
are suddenly released near the surface and start to be observed at this moment.
This sudden release of photons in the shock wave near the stellar surface is
called shock breakout.

The shock breakout signals are strongly affected by the properties of
the SN progenitor and can be used to estimate them.
Especially the duration of the shock breakout signals is
assumed to approximately correspond to the light-crossing time $t_\lc$ of the
progenitor, that is,
$t_\lc \simeq R_\star/c$, where $R_\star$
is the progenitor radius and $c$ is the speed of light.
If a Wolf-Rayet (WR) star explodes, its shock breakout signal
is expected to have a duration of about $1-10$~sec
because typical radii of WR stars are on the order of $1-10$~$R_\odot$
\citep[e.g.,][]{crowther2007,yoon2010}.

The shock breakout durations are, however, suggested
to be extended for several reasons.
For example, if the circumstellar medium (CSM) density is high enough to be optically thick, 
shock breakout durations can be extended as a result of photon diffusion in the
dense CSM \citep[e.g.,][]{balberg2011,svirski2014,svirski2014b}.
Aspherical photospheres may also cause
an extension of the shock breakout durations \citep{suzuki2010,couch2011}.
In addition,
the duration of the shock breakout signals from Type~Ib SN~2008D was
observed to last for $\sim 300$ sec
with a rise time of $\sim 60$ sec \citep{soderberg2008,modjaz2009},
indicating a progenitor radius of $R_\star\simeq t_\lc c \sim 130\ R_\odot$.
The indicated progenitor radius is too large to correspond to a WR star, but
its SN spectral type and
the early light curve modeling point to
a WR progenitor with a radius smaller than 10~$R_\odot$
(e.g., \citealt{chevalier2008,rabinak2011}, see also \citealt{dessart2011,bersten2013}).
The observed long shock breakout duration may also indicate the existence
of mechanisms to extend shock breakout signals.

In this Letter, we argue that shock breakout rise times can be
extended significantly if SN progenitors have inflated envelopes.
In that case, the shock breakout signals diffuse in
the inflated envelope, resulting in the extension of the shock breakout
rise times.
The diffusion times $t_\di$ of canonical WR SN progenitors are
shorter than the light-crossing time $t_\lc$ (Table~\ref{table1}).
However, we show that the diffusion time can be much longer than the light-crossing time
in inflated WR SN progenitors even if their radii are on the order of
the solar radius.

We first explain the inflation of stellar envelopes and
our inflated stellar evolution models in Sect. \ref{inflation}.
We show the effect of the inflated
envelopes on the SN shock breakout signals in Sect. \ref{breakout}.
Our conclusions and their implications are summarized
in Sect. \ref{conclusions}.

\section{Inflation of stellar envelopes}\label{inflation}
We first discuss the inflation of stellar envelopes.
Inflation refers to the extended low-density envelopes
in stellar models that reach Eddington luminosity in their outer layers
\citep{ishii1999,petrovic2006,grafener2012}.
The phenomenon was first discussed by \citet{kato1985,kato1986}
in the context of supermassive stars, considering electron-scattering
as the only opacity source.
The envelope inflation in the models discussed below is, however,
caused by the Fe-group opacity bump at $T\sim 200$ kK 
\citep{ishii1999,petrovic2006,grafener2012}.
When the stellar envelope approaches the Eddington luminosity,
it will become convectively unstable \citep{langer1997}.
The convective luminosity would need to carry any super-Eddington flux
so that the radiative luminosity
does not exceed the Eddington luminosity. However, if the envelope
density is too low, convection cannot become efficient enough
to transport sufficient energy inside the envelope.
Then, the star expands its envelope in order to find a structure with
a lower opacity so that the Eddington luminosity is not exceeded.

To discuss the actual structure of inflated envelopes in WR SN progenitors,
we performed stellar evolution calculations by using
an advanced one-dimensional hydrodynamic binary stellar evolution code
(BEC; see \citealt{heger2000,yoon2006,yoon2010,brott2011})
to model WR stars as nonrotating hydrogen-free helium stars.
The code solves the Lagrangian version of the equations for mass, energy, and
momentum conservation together with convective and radiative energy
transport and is explicitly coupled to a diffusive mixing scheme and 
a nucleosynthesis network (see \citealt{kozyreva2014} for details).
BEC incorporates the latest input physics
\citep{heger2000,yoon2006,yoon2010,brott2011,koehler2014}
and is well-suited to
investigate stars evolving close to the Eddington limit with
inflated envelopes (\citealt{koehler2014}; D. Sanyal et al. in preparation).

\begin{figure}
\centering
\includegraphics[width=\columnwidth]{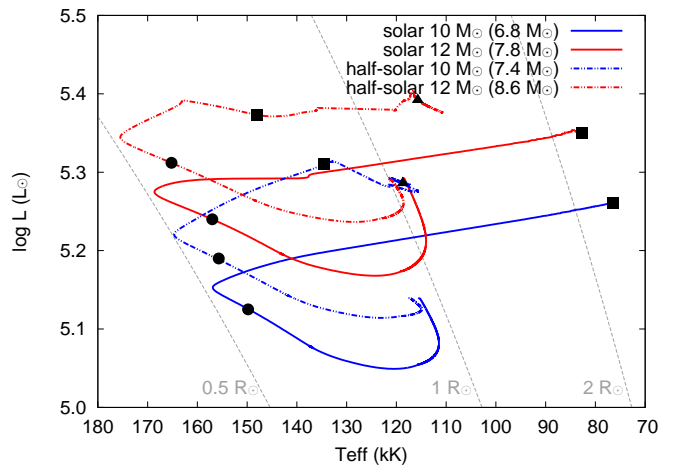}
\caption{
Evolutionary tracks of the helium star models in the
HR diagram. The initial metallicity, initial mass, and final mass of the
 models are shown. The locations of helium-shell
 ignition (circles), core carbon ignition (squares), and core oxygen ignition (triangles) are
 marked.
Lines of the constant radii of 0.5, 1, and 2 \Rsun\ are also shown.
}
\label{HRD}
\end{figure}

In particular, the opacities used to compute the models were
interpolated from the OPAL tables \citep{iglesias1996},
which contain an opacity enhancement around $T\sim 200$ kK due to Fe-group elements.
Convection was treated within the framework of the standard
mixing-length theory \citep{boehm-vitense1958} with the
mixing length set to 1.5 times the local pressure scale height.
The WR mass-loss recipe proposed by \citet{nugis2000} was used for our computations.
The stellar surface was set at the photosphere where the optical depth is 2/3.
We investigate the evolution of helium stars with initial masses of
10 and 12 \Msun\ in this Letter.
We explored two sets of models, one with solar metallicity and the other with half-solar metallicity.
The half-solar metallicity models were computed until the oxygen-burning stage, while solar metallicity models were only computed until carbon ignition
because we encountered numerical problems to iteratively converge models with large
inflated envelopes that are larger at higher metallicity
\citep{ishii1999}.
The inflation in these models at the time
of the SN explosion could thus be somewhat stronger than we assume
here.
The evolutionary tracks of the models in the
Hertzsprung-Russell (HR) diagram are presented in Fig.~\ref{HRD}.
The final stellar properties are summarized in Table~\ref{table1}.

\begin{figure}
\centering
\includegraphics[width=\columnwidth]{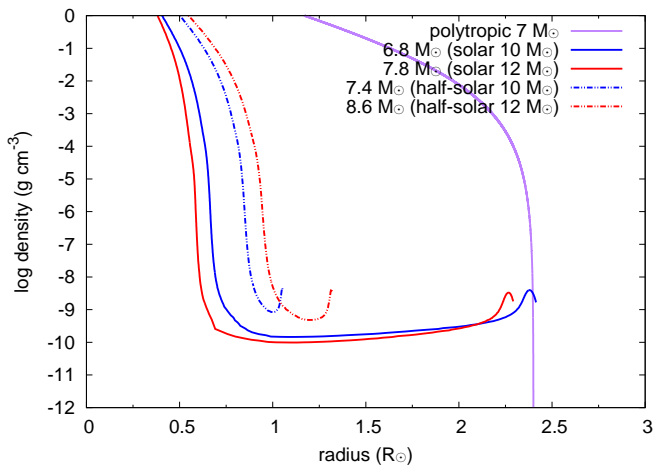}
\caption{
Final density structure of our stellar models.
We also show the structure of a star without the envelope inflation, which is a polytiropic star of 7 \Msun\ and 2.4 \Rsun\ with the
 polytropic index of 3.
The inflated stellar models
have the extended low-density regions on top of the core structure.
}
\label{density}
\end{figure}

The helium zero-age main-sequence models of both 10 and 12 \Msun\ stars are marginally inflated. As a consequence of the applied
mass loss, they lose mass and decrease in luminosity during the core helium-burning phase.
After helium is exhausted in the core, the stars begin to contract and become hotter and
brighter. The models are hardly inflated during this phase because the
Fe-group opacity bump is only partially contained
inside the stars. When the helium-shell ignites, the evolutionary tracks eventually move toward cooler temperatures
because of the mirror principle \citep[e.g.,][]{kippenhahn2013}.
As the models become cooler,
they become significantly inflated and exhibit a pronounced core-halo
structure, as shown in Fig.~\ref{density}.
The electron-scattering Eddington factors of our models
are shown in Table~\ref{table1}. 
When the full opacity is counted into the Eddington factor,
all models reach the critical value of one.
The envelope inflation is stronger for the solar metallicity models
because the prominence of the Fe-group opacity bump increases with
metallicity, as found in the studies by \citet{ishii1999} and \citet{petrovic2006}.
For comparison, we also show 
a polytropic star with $M_\star = 7\ \Msun$ and $R_\star=2.4\ \Rsun$
obtained with the polytropic index of 3 which does not have an inflated envelope.
Figure~\ref{density} shows that our stellar models have
extended low-density layers on top of the core structure.
This type of the envelope inflation is also found in the
previously reported models with similar final core masses \citep{yoon2012}.
The evolution and final position of our stars in the HR diagram 
are consistent with their models.

\begin{figure}
\centering
\includegraphics[width=\columnwidth]{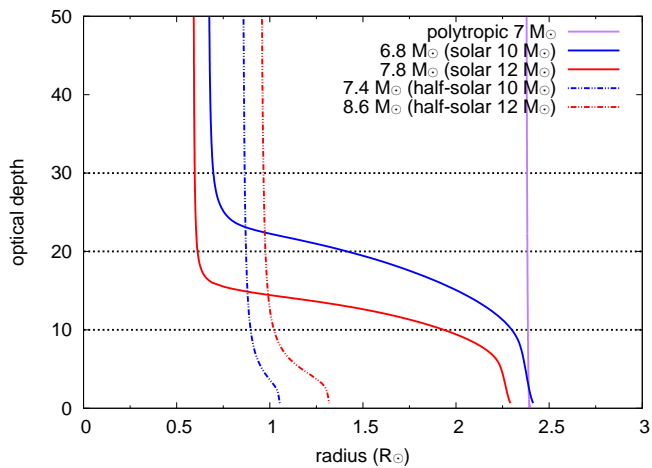}
\caption{
Optical depth $\tau$ in our models.
The optical depth in the polytropic star is obtained with
the constant opacity of 0.2 $\mathrm{cm^2~g^{-1}}$.
}
\label{opticaldepth}
\end{figure}

\begin{table*}
\caption{Stellar properties and corresponding timescales:
Initial metallicity, initial stellar mass, final stellar mass,
final surface electron-scattering Eddington factor,
final mass-loss rate,
final ratio of the mass-loss rate to the critical mass-loss rate,
final stellar radius,
light-crossing time $t_\lc= R_\star /c$,
distance from the radius at $\tau=30$ to $R_\star$,
diffusion time for $\tau=30$ ($v_\sh=10000$ $\mathrm{km~s^{-1}}$),
distance from the radius at $\tau=20$ to $R_\star$,
diffusion time for $\tau=20$ ($v_\sh=15000$ $\mathrm{km~s^{-1}}$),
distance from the radius at $\tau=10$ to $R_\star$, and
diffusion time for $\tau=10$ ($v_\sh=30000$ $\mathrm{km~s^{-1}}$).
} 
\label{table1}      
\centering          
\begin{tabular}{c c c c c c c c c c c c c c}
\hline\hline       
$Z$ &$M_\mathrm{i}$ & $M_\star$ & $\Gamma_e$
& $\dot{M}$ & $\dot{M}/\dot{M}_\mathrm{max}$ & 
$R_\star$   &  $t_\lc$& $\Delta R^{30}$ & $t_\di^{30}$ 
& $\Delta R^{20}$ & $t_\di^{20}$ & $\Delta R^{10}$ & $t_\di^{10}$\\
($Z_\odot$) & ($M_\odot$)  & ($M_\odot$) & & ($10^{-6}$ $M_\odot$ yr$^{-1}$) &&
($R_\odot$) & (sec) &($R_\odot$)  & (sec)&($R_\odot$)  & (sec)&($R_\odot$)  & (sec)  \\
\hline  
  1 & 10 & 6.8 & 0.412 & 8.1 & 0.45  & 2.43  & 5.67 & 1.73 & 121  
                                     &  1.01 & 47.0 & 0.130 & 3.01\\
  1 & 12 & 7.8 & 0.439 & 7.6 & 0.57   & 2.31 & 5.38 & 1.71  & 119 
                                     & 1.70  & 78.6 & 0.372 & 8.62\\
0.5 & 10 & 7.4 & 0.398 & 6.4 & 0.066 & 1.06 & 2.47  & 0.189  & 13.2  
                                     & 0.188  & 8.72 & 0.163 & 3.78\\
0.5 & 12 & 8.6 & 0.428 & 8.3 & 0.11  & 1.33 & 3.10  & 0.356  & 24.9  
                                     & 0.351  & 16.27 & 0.306 & 7.09\\
\hline
- & - & 7$^{(a)}$ & - & - & - & 2.4 & 5.60 & 0.0187 & 1.31  
                                     & 0.0170  & 0.788 & 0.0140 & 0.325\\
\hline
\end{tabular}
\tablefoot{
\tablefoottext{a}{Polytropic star with the polytropic index of 3.}
}
\end{table*}

\section{Shock breakout in inflated stellar envelopes}\label{breakout}
The shock breakout in a SN occurs when photons start to
leak from the shock
wave, that is, when the dynamical timescale $t_\mathrm{dy}$ of the shock
propagation in the remaining unshocked stellar envelope with radius $\Delta R$ ($t_\mathrm{dy}\simeq\Delta R/v_\sh$) becomes
similar to the diffusion timescale
$t_\di\simeq \tau \Delta R/c$,
%\begin{equation}
% t_\di\simeq \frac{\tau \Delta R}{c},
%\end{equation}
where $v_\sh$ is the shock velocity and
$\tau$ is the optical depth in the remaining unshocked stellar envelope
 \citep[e.g.,][]{weaver1976}.
The shock breakout condition can be simply expressed as
$\tau\simeq c/v_\sh$.
%\begin{equation}
% \tau\simeq \frac{c}{v_\sh}.
%\end{equation}
For a shock breakout at the stellar surface without the
inflated envelope, $\Delta R$ is so small that $t_\di$ is smaller
than the light-crossing time $t_\lc \simeq R_\star/c$,
making the shock breakout duration $t_\lc \simeq R_\star/c$.
For example, in the case of the 2.4 \Rsun\ polytropic star we show in Fig.~\ref{density},
the light-crossing time is $t_\lc \simeq 5.60$ sec and the diffusion
time is $t_\di\simeq 1.31$ sec,
assuming $v_\sh=10000$ $\mathrm{km~s^{-1}}$ (Table~\ref{table1}).

The inflated stellar models are characterized by extended
low-density envelopes (Fig.~\ref{density}).
As a result, $\Delta R$ can be much larger than the corresponding values
for the uninflated stellar models.
In Fig.~\ref{opticaldepth}, we show the optical depth in our stellar
models.
In Table~\ref{table1}, we summarize the light-crossing time $t_\lc$ and
the diffusion time $t_\di$ for several shock velocities.
In the uninflated polytropic model, the shock breakout occurs at the edge of the
stellar surface ($\Delta R = 0.0187\ R_\odot$ for $\tau=30$),
making the diffusion time shorter than the light-crossing time.
If the progenitor has an inflated envelope, however, the shock breakout
can occur at the bottom of the inflated envelope, keeping $\Delta R$ large
when the shock breakout occurs.
For example, our solar-metallicity 10 \Msun\ model
has $\Delta R=1.73\ R_\odot$ for $\tau=30$
because of the envelope inflation.
The large $\Delta R$ makes the diffusion time
($t_\di\simeq 121$ sec) much longer
than the light-crossing time ($t_\lc\simeq 5.67$ sec).
Thus,
the timescale of the shock breakout signals within the inflated
envelopes can be dominated by the diffusion time, 
and their rise time is determined by the diffusion time.
The subsequent light curve is expected to decline exponentially
with an $e$-folding timescale of the diffusion time due to 
the photon diffusion in the shocked envelope.
Overall, the total shock breakout duration becomes significantly
longer than the light-crossing time.
The dynamical timescales of our inflated envelopes
at the time of their explosions
are $\sim 2000$ sec, and the inflated envelopes can be sustained
until the shock passes through.

We have computed the diffusion times for several shock velocities
between 10000 and 30000 $\mathrm{km~s^{-1}}$.
If the shock velocity at the shock breakout is too high, the diffusion
time becomes shorter than the light-crossing time. In our models,
the diffusion time starts to be shorter than the light-crossing time when
the shock velocity is higher than $\sim 30000$ $\mathrm{km~s^{-1}}$ (Table~\ref{table1}).
The shock velocity is suggested to reach above 30000 $\mathrm{km~s^{-1}}$
at the shock breakout in WR stars \citep[e.g.,][]{nakar2010}.
However, previous studies assume that WR stars only have a steeply
declining envelope as in our polytropic model.
A flat inflated envelope exists on top of the steeply declining core
in the density structure of our models.
The flat density structure decelerates the shock significantly, and
the shock velocity at the shock breakout in inflated envelopes
can be lower than 30000 $\mathrm{km~s^{-1}}$.
Because the shock propagation in inflated stellar structure has not yet been
investigated, hydrodynamical modeling in inflated envelopes is
required to estimate the shock velocity at breakout in this case.

The sizes of the inflated
envelopes in the half-solar-metallicity models are smaller than those in the solar-metallicity models
because the Fe-group opacity bump is weaker with decreasing metallicity.
However, the diffusion time can
still be longer than the light-crossing time, and the shock breakout
rise times can be longer than 10 sec in these WR stars.

Although higher metallicity models are preferred to have large
$\Delta R$ and thus long $t_\di$ in terms of opacity, the inflated envelope
can disappear because of higher mass-loss rates.
\citet{petrovic2006} showed that the inflated envelope disappears when
the mass-loss rate of the star is higher than the critical mass-loss rate
defined as
\begin{equation}
\dot{M}_\mathrm{max}\simeq 4\pi R_m^2 \rho_m\sqrt{\frac{G M_\star}{R_m}},
\end{equation}
where $\rho_m$ is the lowest density in the inflated envelope
and $R_m$ is the radius at the lowest density \citep[see also][]{grafener2012}.
We show the ratio of the mass-loss rate in our models to the critical
mass-loss rate in Table~\ref{table1}. 
Our solar-metallicity models are close to the critical mass-loss rates.
Higher-metallicity stars are likely to exceed the critical mass-loss rate.

Although we have focused on WR SN progenitors in this Letter,
stars do not need to be WR stars to have inflated envelopes.
When stars are near the Eddington luminosity, they can have inflated
envelopes even if they are hydrogen-rich.
For example, luminous blue variables are hydrogen-rich stars that are
close to the Eddington luminosity, and they have recently been suggested to be
SN progenitors \citep[e.g.,][]{gal-yam2009}.
They can also have inflated envelopes when they explode.

To confirm that some SN progenitors are actually inflated stars,
observational consequences of the inflated envelopes in SN
observational properties other than the shock breakout rise times
are required to be investigated.
Because
the inflated envelopes in our models only have $\sim 10^{-8}$ \Msun, they are not likely
to strongly affect later SN observables such as its light curve.
However, the shock breakout spectra or very early SN spectra such as that recently
obtained by \citet{gal-yam2014} may be affected by an inflated envelope.
It is especially important to distinguish the shock breakout 
extension caused by an inflated envelope from that produced by a dense
stellar wind because the shock breakout signals can be extended by the
photon diffusion in both cases. 
Estimates of the progenitor mass-loss rate, for instance from multiwavelangth
SN observations, may constrain the wind density near the
stellar photosphere and thereby confirm or rule out a shock break-out extension
caused by winds (see Sect.\,\ref{sn2008D} for the case of SN\,2008D).
\citet{maeda2013} also discusses possible observational consequences
of the inflated envelopes.

\subsection{SN 2008D}\label{sn2008D}
The extension of shock breakout signals by an inflated envelope
may explain the long duration of
the shock breakout signal observed in SN~2008D.
Even if the progenitor of SN~2008D is a WR star with $R_\star\sim R_\odot$,
our result show that the rise time of the shock breakout signals can be $\sim 60$ sec 
because of the long diffusion time in the inflated envelope
below which the shock breakout occurs.
The shock-breakout luminosity decline after the luminosity peak is found to be
similar to an exponential decay for a while \citep{soderberg2008},
which is consistent with the diffusion interpretation.

The observed high temperature of the shock breakout signals in SN~2008D
($\sim 0.1-1$ keV, e.g., \citealt{modjaz2009,li2008})
is suggested to indicate the shock velocity
above $\sim 20000$ $\mathrm{km~s^{-1}}$ 
\citep[e.g.,][]{katz2010,balberg2011}.
The longest diffusion time
for $v_\sh=20000$ $\mathrm{km~s^{-1}}$ among
our models is 19 sec from the solar 12
\Msun\ model.
Although it is shorter than observed, $\sim60$ sec,
the diffusion time is still much longer than the light-crossing time.
The observed $e$-folding timescale in the declining phase
($\sim 130$ sec, \citealt{soderberg2008}) is also longer than
the diffusion time in our models, but our models do predict the overall
extension of the shock breakout signals.
The shock temperature becomes
$\sim 0.1$ keV using $v_\sh=20000$ $\mathrm{km~s^{-1}}$ and 
our inflated envelope density \citep{katz2010}, which
corresponds to the observational constraints \citep[e.g.,][]{modjaz2009}
and can explain the observed thermal X-ray luminosities
($\sim 10^{43}$ $\mathrm{erg~s^{-1}}$) with a radius of $\sim \Rsun$.
More investigations are required to see whether
the X-ray spectra can be explained by our model.

The later shock velocity is observed to be $0.25c$ \citep{soderberg2008}.
However, the shock can have accelerated after the shock breakout because of
the steep density decline between the stellar surface and the low
density wind (cf., \citealt{tolstov2013}). Thus, the observed velocity
may not correspond to the actual shock velocity at the shock breakout.

The estimated SN ejecta mass of SN~2008D is $3-7$ \Msun\
\citep{soderberg2008,mazzali2008,tanaka2009,bersten2013}.
Our pre-SN stellar models (Table~\ref{table1}) have $6.3-7.8$ \Msun.
Subtracting a remnant mass of about 1.4 \Msun, the expected ejecta
masses from our inflated stellar models are consistent
with those of SN~2008D.
The mass-loss rates of our models are also consistent with that
estimated from radio observations
($7\times 10^{-6}\ M_\odot~\mathrm{yr^{-1}}$, \citealt{soderberg2008}).

The metallicity at the location where SN~2008D appeared is estimated to
be 0.5~$Z_\odot$ based on 
the metallicity gradient in the host galaxy \citep{soderberg2008}.
Our 0.5~$Z_\odot$ stars do not inflate enough to explain the long shock
breakout rise time of $\sim 60$ sec.
However, Galactic metallicity gradients have large dispersions
\citep[e.g.,][]{pedicelli2009}.
Since our $Z_\odot$ stars have much longer diffusion times,
it is possible that the progenitor metallicity of SN~2008D is
around $Z_\odot$ or higher,
and the progenitor had a sufficiently inflated envelope to show
the long shock breakout rise time with the estimated shock
velocity.
Recent metallicity measurements at the location of SN~2008D
support our arguments \citep{modjaz2011,thone2009}.

\section{Conclusions}\label{conclusions}
We have shown that the rise times of supernova shock breakout signals 
can be extended because of inflated stellar envelopes.
The shock breakout can occur within the low-density inflated envelopes
in which the shock breakout signals are diffused and extended.
The long diffusion time in the inflated envelopes makes the shock
breakout rise times long. The shock breakout timescale is then dominated by
the diffusion time instead of by the light-crossing time.
Even if a SN progenitor has a radius on the order of the solar radius
whose light-crossing time is a few seconds,
the rise time of the shock breakout signals can be more than 100 seconds
because of the inflated envelope.
The extension of the shock breakout
signals by the inflated envelope may explain the mysterious long shock breakout
rise time observed in Type~Ib SN~2008D,
although more investigations are required to confirm this.

The actual existence of inflated stellar envelopes is still widely debated. 
While we find the envelope inflation in our stellar evolution models,
the envelope inflation is not as prominent in other models
\citep[e.g.,][]{ekstrom2012}.
If inflated envelopes exist, they are also
pulsationally unstable \citep[e.g.,][]{glatzel1993}
with unknown consequences.
If we can confirm that SN shock breakout signals
from WR SN progenitors have signatures of inflated envelopes,
for example, extended shock breakout signals from SN progenitors with low 
mass-loss rates,
it may indicate the common existence of inflated stellar envelopes 
in nature. The shock breakout signals can be a viable
probe of unsolved problems in stellar structure as well as a touchstone
of the stellar evolution theory, which predicts the final status of
the massive stars.

\begin{acknowledgements}
TJM is supported by Japan Society for the Promotion of
 Science Postdoctoral Fellowships for Research Abroad
 (26\textperiodcentered 51).
\end{acknowledgements}

\end{document}